\documentclass{iopart}
\usepackage{iopams}  

\usepackage{graphicx}
\usepackage[caption=false]{caption}
\usepackage{subfig}
\usepackage{hyperref}
\usepackage{xspace}
\usepackage{wrapfig}

\newcommand{\epair}{\ensuremath{e^+e^-} pair\xspace}
\newcommand{\epairs}{\ensuremath{e^+e^-} pairs\xspace}
\newcommand{\pion}{\ensuremath{\pi^0}\xspace}
\newcommand{\fig}[1]{{Fig.~\ref{#1}}\xspace}
\newcommand{\eg}{\emph{e.\,g.}\xspace}%
\newcommand{\pp}{{\ensuremath{p+p}}\xspace}
\newcommand{\AuAu}{\ensuremath{\mathrm{Au}+\mathrm{Au}}\xspace}
\newcommand{\sqrts}{\ensuremath{\sqrt{s}}\xspace}
\newcommand{\sqrtsnn}{\ensuremath{\sqrt{s_{NN}}}\xspace}
\newcommand{\pT}{\ensuremath{p_{\mathrm T}}\xspace}
\newcommand{\gevc}{GeV/\ensuremath{c}\xspace}
\newcommand{\kevcc}{keV/\ensuremath{c^2}\xspace}
\newcommand{\mevcc}{MeV/\ensuremath{c^2}\xspace}

\begin{document}

\title[Thermal photons in heavy ion collisions with PHENIX]{Measurements of thermal photons in heavy ion collisions with PHENIX}

\author{T Dahms (for the PHENIX Collaboration)}

\address{Department of Physics and Astronomy,
Stony Brook University,
Stony Brook, NY 11794-3800, USA}
\ead{torsten.dahms@stonybrook.edu}
\begin{abstract}
  Thermal photons are thought to be the ideal probe to measure the
  temperature of the quark-gluon plasma created in heavy ion
  collisions. PHENIX has measured direct photons with $\pT < 5$~\gevc
  via their internal conversions into \epairs in \AuAu collisions at
  \sqrtsnn = 200~GeV and has now provided a baseline measurement from
  \pp data.
\end{abstract}

%\noindent{\it Keywords}: Thermal photons, Quark-gluon plasma
%\submitto{\jpg}
%\maketitle

\section{Introduction}
\label{sec:introduction}
Direct photons are produced during all stages of heavy ion collisions
at the Relativistic Heavy Ion Collider (RHIC). Because they do not
interact strongly, they escape the medium unaffected by final state
interactions and provide a promising signature of the earliest and
hottest stage of the quark-gluon plasma
(QGP)~\cite{PhysRevC.69.014903}. The main sources of direct photons
from a QGP are quark-gluon Compton scattering ($q g \rightarrow \gamma
q$), quark-antiquark annihilation ($q \bar{q} \rightarrow \gamma g$)
and brems\-strahlung involving thermalized and incoming
partons~\cite{PhysRevD.58.085003}. At RHIC energies thermal photons
are predicted to be the dominant source of direct photons in $1 < \pT
< 3$~\gevc~\cite{PhysRevC.69.014903}.

The formation of a hot and dense medium has been established by
experimental results at RHIC~\cite{Adcox2005}. The measurement with
the Electromagnetic Calorimeter (EMCal), based on a statistical
subtraction of the expected background form hadronic sources, remains
limited at low \pT due to large systematic
uncertainties~\cite{adler:012002,adler:232301}.

The uncertainties in the knowledge of the decay background can be
reduced by avoiding its explicit measurement. A tagging method has
been developed in which a very pure photon sample is selected with
strict photon identification cuts~\cite{Gong:2007hr}. In this sample
those photons are tagged, which combined with other clusters in the
EMC (selected with less stringent photon identification cuts) can be
identified via their invariant mass as the result of a \pion decay. In
the ratio of pure photons and tagged photons the reconstruction
efficiency of the pure photons cancel and with it the associated
systematic uncertainty.

Furthermore, to circumvent the limitations due to the energy
resolution at low photon energies, the excellent capabilities of the
PHENIX detector to measure electrons can be utilized by measuring
photons via their external conversion into
\epairs~\cite{Dahms:2007rs}. These conversion pairs can be used as an
alternative clean photon sample in the method described above.

\section{Internal Conversions}

To overcome the statistical and systematic limitations an alternative
approach has been pursued~\cite{ppg086} and is outlined in the
following. Any process that produces a real photon can also create a
virtual photon which converts internally into an \epair. Therefore,
\epairs are also produced through internal conversions of virtual
direct photons, \eg via the $q g$ Compton scattering ($q g \rightarrow
\gamma^* g \rightarrow e^+ e^- g$). The relation between photon
production and the associated \epair production can be written as:
\begin{equation}\label{eq:kroll-wada}
\frac{d^2n_{ee}}{dm} = \frac{2 \alpha}{3 \pi} \frac{1}{m} \sqrt{1 -
  \frac{4 m_e^2}{m^2}} \left(1+\frac{2m_e^2}{m^2}\right) S dn_{\gamma}
\end{equation}
with $m$ being the mass of the \epair and $m_e=511$~\kevcc the mass of
the electron. The process dependent factor $S$ goes to 1 as $m
\rightarrow 0$ or $m \ll \pT$. For \pion and $\eta$ decays, $S$ is
given by $S = |F(m^2)|^2 (1-m^2/m_h^2)^3$~\cite{landsberg} where $m_h$
is the hadron mass and $F(m^2)$ the form factor. For \epair masses
approaching $m_h$, the factor $S$ goes to zero. While the measurement
of real thermal photons suffers from a large background of hadron
decays, measuring virtual photons allows to select a mass range, in
which \pion decays are suppressed due to the cut-off in $S$.

\begin{figure}[b]
  \centering
  \subfloat{\label{fig:masscomp}\includegraphics[height=0.22\textheight]{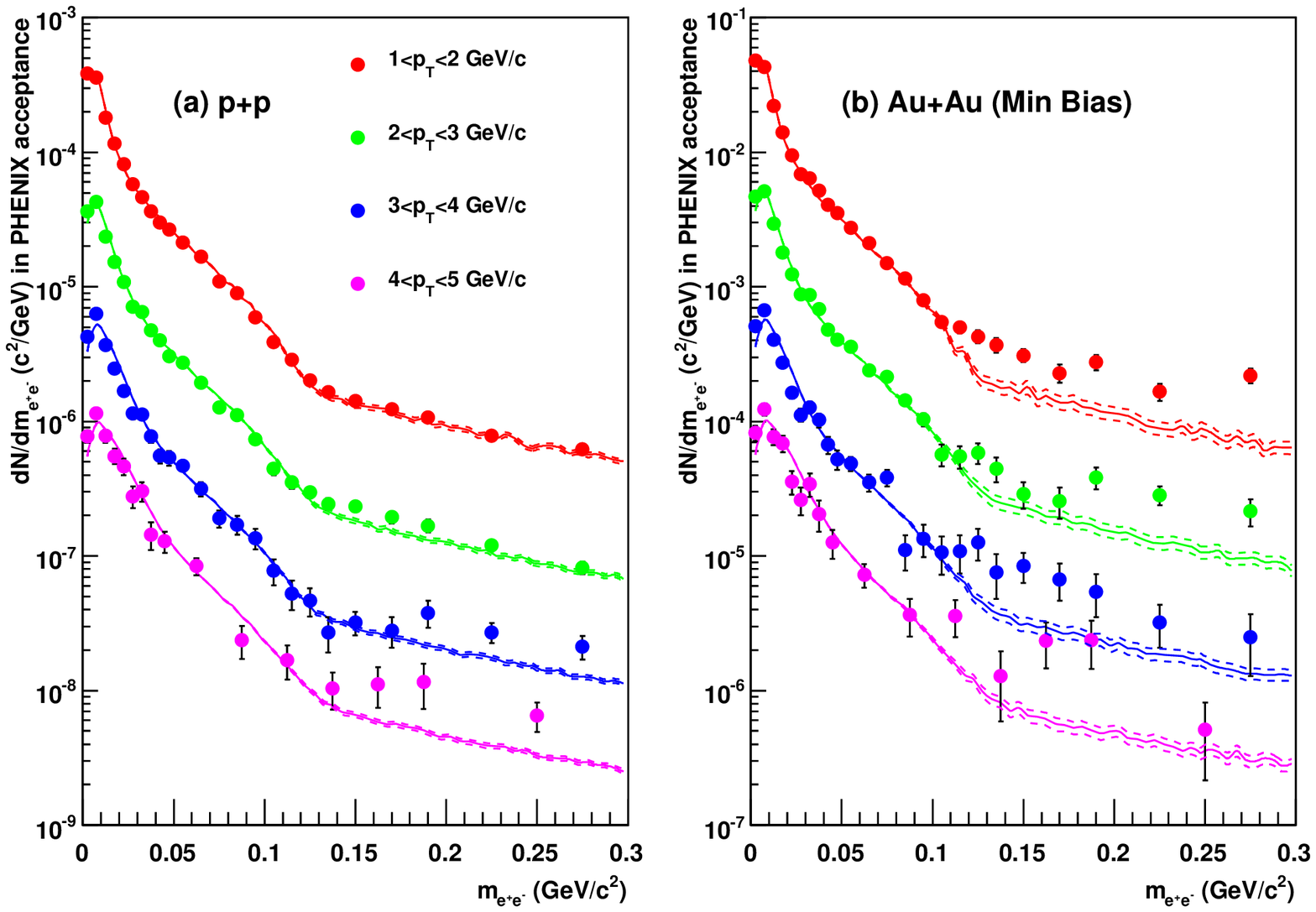}}
  \subfloat{\label{fig:massfit}\includegraphics[height=0.22\textheight]{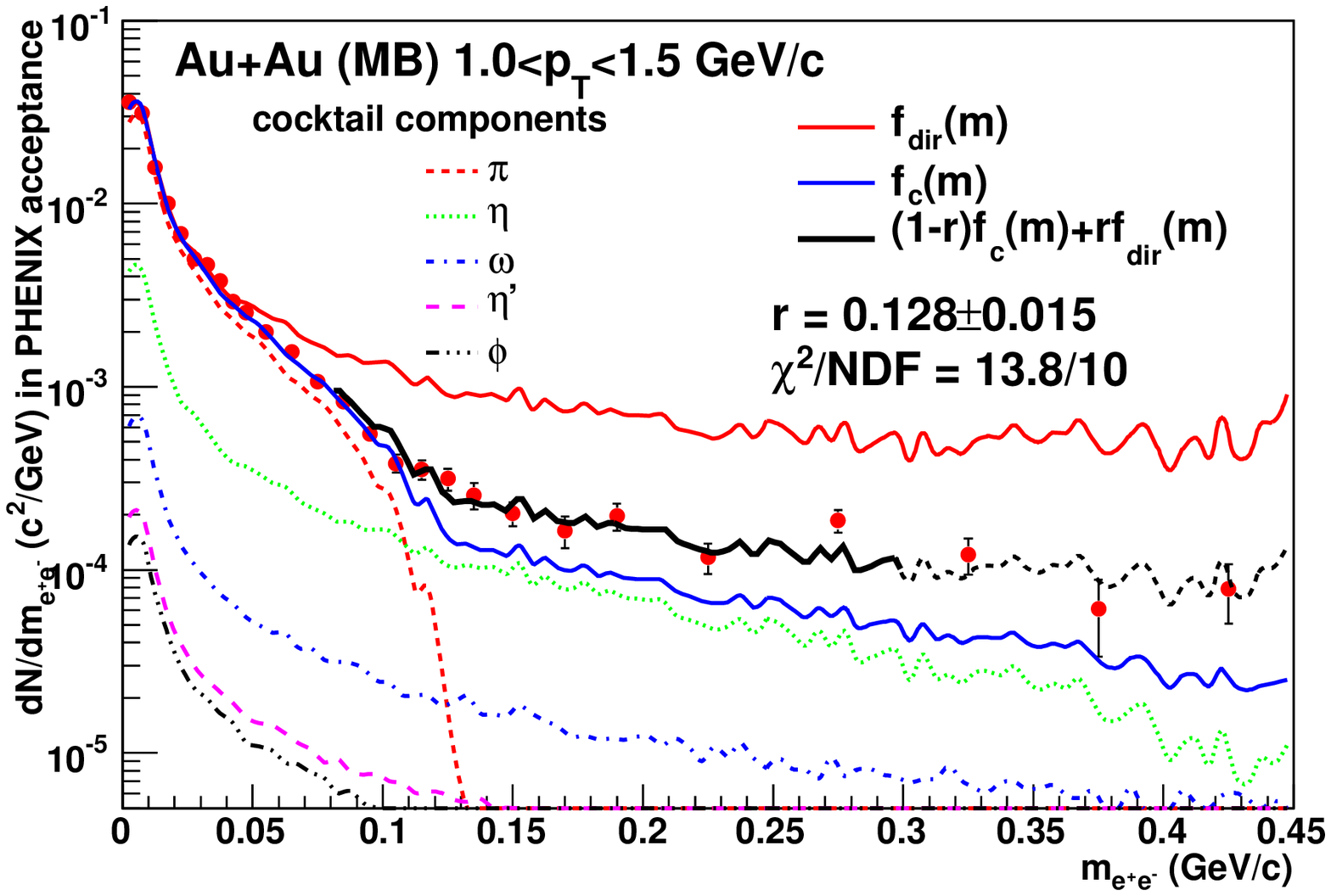}}
  \caption{Left: The \epair invariant mass distributions in (a) \pp
    and (b) minimum bias \AuAu collisions compared to a cocktail of
    hadronic sources. Right: Mass distribution of \epairs for \AuAu
    minimum bias events for $1.0 < \pT < 1.5$~\gevc with a fit as
    described in the text.}
  \label{fig:mass}
\end{figure}

The following analysis is based on 800M minimum bias events collected
during the \AuAu run at \sqrtsnn = 200~GeV in 2004 and 2.25~pb$^{-1}$
of single electron triggered data recorded during the \pp run at
\sqrts = 200~GeV in 2005. Electrons have been identified using
information of the Ring Imaging Cherenkov detector (RICH) and a
matching of momentum and the energy deposited in the EMC. Details of
the analyses can be found in~\cite{ppg085, ppg075}. The invariant mass
spectrum of \epairs after background subtraction is shown in
\fig{fig:mass} for \pp (left) and \AuAu (right) for various \epair-\pT
bins above 1~\gevc. The data are corrected for the electron
identification efficiency based on a full Monte Carlo simulation, but
are not corrected for the geometrical acceptance~\cite{ppg085,
  ppg075}. These data are then compared to a cocktail of hadron decays
from a fast Monte Carlo simulation whose input is based on the
measured hadron spectra~\cite{ppg085, ppg075}. The cocktail is
normalized to the data in the mass range 0--30~\mevcc. While in \pp
the agreement of the cocktail to the data is remarkable and a small
excess is only observed at very high \pT and large mass, the \AuAu
data show a large excess over the full \pT range of 1--5~\gevc.

The observed excess above the hadronic cocktail can be analyzed under
the assumption that it is solely due to internal conversions of direct
virtual photons. While at zero mass \epairs from hadron decays have
the same shape in mass as internal conversions of direct photons, the
suppression due to $S$ when approaching the hadron mass changes the
shape of \epairs from hadron decays. Therefore, one can fit, after an
initial normalization to the mass range 0--30\mevcc, the two expected
shapes for \epairs from hadronic decays $f_{\rm cocktail}$ (as shown
in~\fig{fig:mass}) and from direct photons $f_{\rm direct}$ in the
mass range 80--300\mevcc in which the \pion contribution is severely
suppressed. The only fit parameter is the relative fraction of direct
photons~$r$:
\begin{equation}
f(m) = (1-r) f_{\rm cocktail}(m) + r f_{\rm direct}(m).
\end{equation}

As an example, the fit result is shown in~\fig{fig:mass} for \epairs
with $1.0 < \pT < 1.5$~\gevc in minimum bias \AuAu collisions. The
quality of the fit result (\eg, $\chi^2/NDF = 13.8/10$ for the lowest
\pT bin) indeed justifies the assumption that the observed excess is
due to internal conversion of virtual photons. A fit with the mass
shape of \epairs from $\eta$ Dalitz decays would lead to a two times
larger $\eta$ cross section than measured~\cite{adler:202301} and a
significantly worse $\chi^2/NDF=21.1/10$. Little contribution from
other sources is expected to this kinematic region, as it is limited
to $m < 2 m_{\pi}$. PHENIX has reported a large dielectron enhancement
for $150<m<750$~\mevcc in~\cite{ppg075}, but the dominant fraction of
the enhancement is concentrated at low \pT~\cite{Toia:2008qm}.

\section{Results}
\begin{figure}[b]
  \centering
  \includegraphics[height=0.3\textheight]{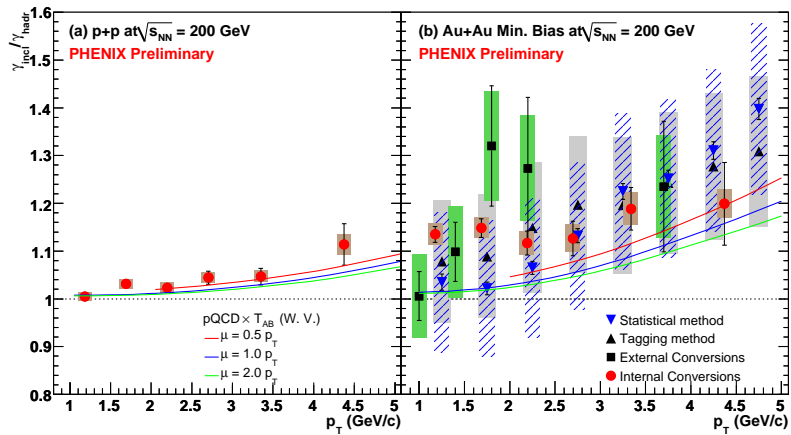}
  \caption{Direct photon excess. (a) The fraction of the direct photon
    component as a function of \pT in \pp. (b) \AuAu (min. bias)
    compared to other measurements of direct photons as described in
    the legend. The curves are from a NLO pQCD
    calculation~\cite{nlo}.}
  \label{fig:ratio}
\end{figure}
The result is shown as $1+r \approx \gamma_{\rm incl}/\gamma_{\rm
  hadr}$ in the left of~\fig{fig:ratio} for \pp collisions. It is
compared to a NLO pQCD calculation of direct photons~\cite{nlo}.
While the fraction of direct photons measured in \pp agrees well with
the pQCD calculation, a clear excess is observed in minimum bias \AuAu
collisions shown in the right of~\fig{fig:ratio}. The \AuAu result is
also compared to the other measurements of direct photons at low \pT;
the conventional statistical subtraction of hadronic decay
photons~\cite{isobe:S1015}, the tagging method~\cite{Gong:2007hr} and
the external conversion analysis~\cite{Dahms:2007rs}. While they all
agree within their uncertainties, the improvement in statistical and
systematic uncertainties is quite significant.

\begin{wrapfigure}{l}{0.46\textwidth}
  \includegraphics[width=0.45\textwidth]{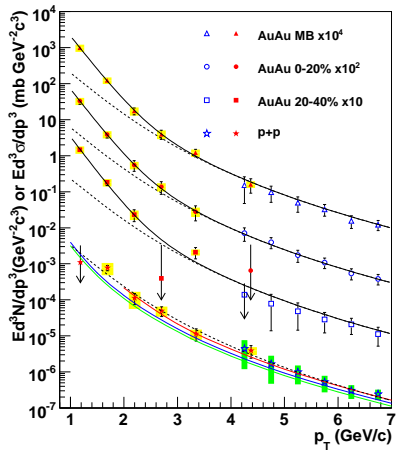}
  \caption{Invariant cross section (\pp) and invariant yield (\AuAu)
    of direct photons. The \pp result is compared to a NLO pQCD
    calculation. Solid points are from this analysis, open points
    from~\cite{adler:012002,adler:232301}. The dashed lines show a
    $T_{AA}$ scaled, modified powerlaw fit of \pp. The black curves
    show an exponential plus the $T_{AA}$ scaled \pp fit.}
  \label{fig:ptspectra}
\end{wrapfigure}

The fraction of direct photons $r$ can be converted into a direct
photon spectrum by multiplying with the inclusive photon spectrum. The
inclusive photon yield is determined for each \pT bin by
$dN_{\gamma}^{\rm incl} = dN_{ee}^{\rm data} \times dN_{\gamma}^{\rm
  cocktail}/dN_{ee}^{\rm cocktail}$, where $dN_{ee}^{\rm data}$ and
$dN_{ee}^{\rm cocktail}$ are the yields of \epairs in $m < 30$~\mevcc
for data and cocktail, respectively, and $dN_{\gamma}^{\rm cocktail}$
is the yield of photons from the cocktail. The resulting direct photon
spectra for \pp and \AuAu collisions are shown
in~\fig{fig:ptspectra}. The \pp data can be well described with a
modified powerlaw function: $A_{pp} (1+\pT^2/b)^{-n}$. The direct
photons measured in \AuAu are above the binary scaled fit to the \pp
result. An exponential fit plus the binary scaled modified power law
($A {\rm e}^{-\pT/T} + T_{AA} \times A_{pp} (1+\pT^2/b)^{-n}$) reveals
an inverse slope of $T = 220 \pm 23 \pm 18$~MeV for central
collisions, suggesting an excess possibly of thermal origin. The
inverse slope of the direct photon spectrum can be related to the
intitial temparature of the QGP~\cite{denterria:451}.

\end{document}